\documentclass{ecai}
\usepackage{graphicx}
\usepackage{latexsym}

\usepackage{booktabs}

\usepackage{changepage}
\usepackage{adjustbox}
\usepackage{multirow}
\usepackage{amsmath}
\usepackage{bm}
\usepackage{algorithm}
\usepackage{algorithmic}
\usepackage{subfigure}
\usepackage[bb=boondox]{mathalfa}
\DeclareMathOperator*{\minimize}{minimize}

\begin{document}

\begin{frontmatter}

\title{User-Controllable Recommendation via Counterfactual Retrospective and Prospective Explanations}

\author[1]{\fnms{Juntao}~\snm{Tan}\thanks{Corresponding Author. Email: juntao.tan@rutgers.edu.}}
\author[1]{\fnms{Yingqiang}~\snm{Ge}}
\author[2]{\fnms{Yan}~\snm{Zhu}} 
\author[2]{\fnms{Yinglong}~\snm{Xia}} 
\author[3]{\fnms{Jiebo}~\snm{Luo}} 
\author[1]{\fnms{Jianchao}~\snm{Ji}} 
\author[1]{\fnms{Yongfeng}~\snm{Zhang}}

\address[1]{Rutgers University} 
\address[2]{Meta Platforms, Inc.}
\address[3]{University of Rochester}

\address[]{\{juntao.tan, yingqiang.ge\}@rutgers.edu, yzhuuu@gmail.com, yxia@fb.com, jluo@cs.rochester.edu, \\\{jianchao.ji, yongfeng.zhang\}@rutgers.edu} 

\begin{abstract}
Modern recommender systems utilize users' historical behaviors to generate personalized recommendations. However, these systems often lack user controllability, leading to diminished user satisfaction and trust in the systems. Acknowledging the recent advancements in explainable recommender systems that enhance users' understanding of recommendation mechanisms, we propose leveraging these advancements to improve user controllability. In this paper, we present a user-controllable recommender system that seamlessly integrates explainability and controllability within a unified framework. By providing both retrospective and prospective explanations through counterfactual reasoning, users can customize their control over the system by interacting with these explanations.

Furthermore, we introduce and assess two attributes of controllability in recommendation systems: the complexity of controllability and the accuracy of controllability. Experimental evaluations on MovieLens and Yelp datasets substantiate the effectiveness of our proposed framework. Additionally, our experiments demonstrate that offering users control options can potentially enhance recommendation accuracy in the future. Source code and data are available at \url{https://github.com/chrisjtan/ucr}.

\end{abstract}

\end{frontmatter}

\section{Introduction}
\begin{figure}[t]
    \centering
    \includegraphics[width=0.9\linewidth]{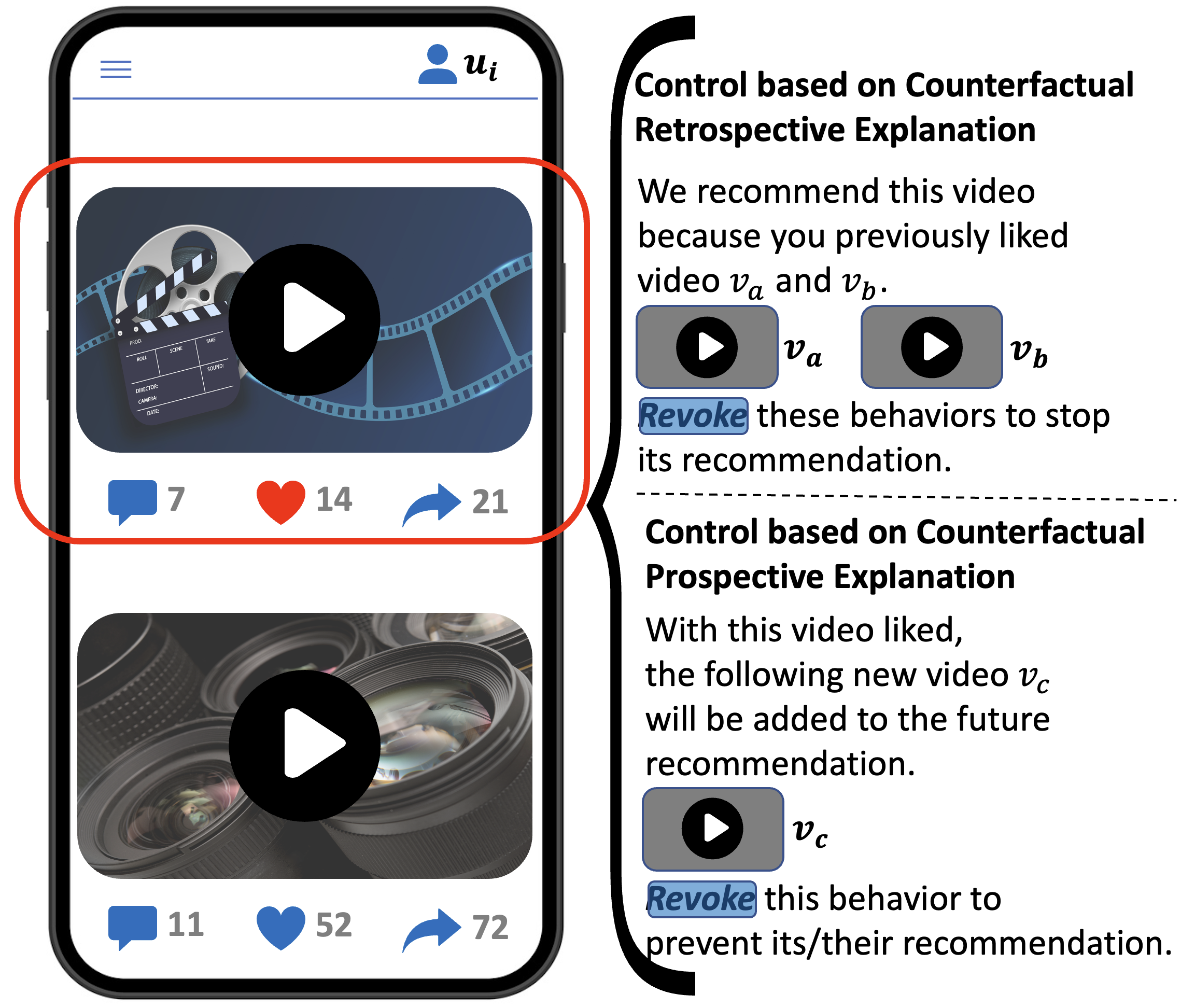}
    \caption{A toy example illustrates the concept of user-controllable recommendations. For the recommended video highlighted by a red bounding box, the retrospective explanation (displayed in the top-right) informs users of the past behaviors that resulted in the video's recommendation, while the prospective explanation (shown in the bottom-right) conveys the impact of liking this video. Users have the option to click the ``revoke'' button to rescind related behaviors.}
    \label{fig:overview}
\end{figure}
Controllability is crucial in developing trustworthy recommender systems. Studies have demonstrated that offering users the ability to control recommender systems can enhance satisfaction and trust in the recommendation results \cite{knijnenburg2012inspectabilityandcontrol, xiao2007commerce, Jannach2019explanationsandcontrol, jugovac2017interacting, ge2022survey}. However, current recommender systems typically offer limited user control, leaving users to passively receive and accept the recommendations provided. If users are dissatisfied with the suggested items, there is often no way for them to intervene, adjust, or modify the recommendation lists in a better direction. Moreover, users are uncertain about what the recommendation models have learned about them and based on what information. Such uncertainties can raise privacy concerns and further impair users' trust in recommender systems.

Therefore, some research has sought to increase controllability in recommendations by enabling users to explicitly express their preferences on pre-defined aspects or features, which is a common approach in the industry \cite{knijnenburg2011each, hijikata2012relation, wasinger2013scrutable,balog2019transparent}. However, users may not always be able to accurately address their preferences \cite{jannach2016usercontrol}, and their preferences may evolve over time, rendering past preferences outdated \cite{hu2008collaborativefiltering,Amatriain2009rateitagain, jannach2016usercontrol, Jannach2019explanationsandcontrol}. Besides, due to the opaque nature of the recommendation mechanism, users are not directly aware of the effects or consequences of their expressed preferences \cite{zhang2020explainable,tan2021counterfactual}. Alternative approaches allow users to implicitly update their preferences by interacting with the recommendation results \cite{Harper2015putting, lamche2014interactive}. In these studies, the lack of understanding remains a primary concern, as users are not informed about how their interactions update their preferences or what the implications of their feedback are in the future.

Upon examining the aforementioned research, we notice that the challenge of achieving optimal user controllability comes from the inherent opacity and black-box nature of recommendation systems. Users are often unaware of how their past behaviors influence present recommendations or how their current behavior will affect future recommendations. Allowing users to manipulate AI systems without a comprehensive understanding of the underlying mechanisms may lead to more harm than benefit \cite{barbosa2020you}. Achieving effective controllability extends beyond simply offering an interface. A user-controllable recommender system should be capable of explaining the reasoning behind its recommendations and the consequences of each user action. Therefore, we believe that harnessing the explainability of these systems has the potential to enhance user controllability.

In this paper, we propose a user-controllable recommendation framework with two essential properties. The first property, \textbf{retrospective controllability} with \textbf{retrospective explanation}, enables users to identify how their previous behaviors, such as clicks, purchases, and likes, contribute to current recommendations. The system should also provide the ability to delete or disregard specific past behaviors. This concept aligns with users' \textit{right to be forgotten (RTBF)} on the internet \cite{art17gdpr}. The second property, \textbf{prospective controllability} with \textbf{prospective explanation}, offers users insight into how their interaction with the currently recommended item(s) will influence future recommendations. By operating in an informed manner, users can effectively control their future recommendations toward desired directions.

To achieve these objectives, we investigate the integration of explainability and controllability in a unified recommender system. Recent advancements in counterfactual reasoning have enabled accurate explanations for black-box machine learning models \cite{stepin2021survey, tan2021counterfactual, goyal2019counterfactual, tan2022learning, tan2023explainablefold, mothilal2020explaining, ge2022explainable, ghazimatin2020prince}. The core concept of counterfactual reasoning aligns well with user-controllable recommendation goals. Both retrospective and prospective controllability involve addressing counterfactual "what if" questions: 1) What recommendations would have been made if the user had (not) interacted with specific items previously? and 2) How will future recommendations change if the user interacts with certain items presently? Providing counterfactual retrospective and prospective explanations empowers users with a comprehensive understanding of the consequences of their past and current behaviors, allowing them to selectively exclude such behaviors when necessary. Thus, we propose a model-agnostic framework that generates counterfactual retrospective and prospective explanations, informing users of their past and future recommendation impact and enabling informed decision-making and intervention in the recommendation process through engagement with these explanations.

Another significant contribution of this research is the incorporation of a quantitative evaluation of controllability in recommender systems, an aspect often overlooked in previous controllable recommendation studies. We introduce distinct evaluation protocols for retrospective and prospective controllability. For retrospective controllability, we assess the framework from two unique perspectives: 1) complexity of controllability, which quantifies the cost of intervening in past behaviors to modify current recommendations, and 2) accuracy of controllability, measuring the system's effectiveness in adjusting recommendations according to provided explanations when users implement suggested modifications. In terms of prospective controllability, we simulate user behaviors in a real-world context and demonstrate that the inclusion of control options has the potential to yield more accurate recommendations in the future. This assertion is supported by standard evaluation metrics in recommendation.

In summary, our work contributes the following:
\begin{itemize}
\setlength\itemsep{0pt}
\item We propose the first user-controllable recommendation framework utilizing counterfactual explanations.
\item We define two categories of counterfactual explanations for user-controllable recommendations: counterfactual retrospective explanations and counterfactual prospective explanations.
\item We introduce two perspectives for evaluating controllability and, based on these, develop two metrics to enable the standardized assessment of controllable recommendation systems.
\item For the first time, we quantitatively demonstrate the potential for improving recommender system accuracy by providing users with additional control options.
\end{itemize}

\section{Related Works}
\label{sec:related}
Most of the existing research on controllable recommendation is conducted from a user study perspective rather than a machine learning perspective. Early works in controllable recommendation focus on explicitly defining users' preferences regarding specific content. For instance, in \cite{knijnenburg2011each}, an interaction strategy was introduced where users can indicate the weights they assign to each item attribute to express their preferences. Similarly, \cite{hijikata2012relation} implemented a music recommender system that enables users to intervene in the recommendation process by selecting preferred item categories or editing their profiles. The authors observed that when the system provides highly precise recommendations, user intervention leads to increased satisfaction. In \cite{wasinger2013scrutable}, a recommender system was proposed for recommending menus at restaurants, allowing users to update their preferences and provide information about allergies towards food ingredients. However, these methods necessitate users to explicitly answer questions, which may not always be feasible as users might face difficulty accurately expressing their preferences.

Some other studies incorporate implicit updating of user preferences through dynamic feedback. In these studies, the terms ``interactive'' and ``controllable'' are often used interchangeably. In \cite{Harper2015putting}, users were given the ability to directly interact with the recommendation results by re-ranking the generated recommendation lists to align with their expectations. Upon re-ranking, the system automatically updates the users' preferences for two attributes: popularity and age of the items. However, \cite{Harper2015putting} has certain limitations, such as the inability for users to remove disliked items and a limited number of item attributes. \cite{bostandjiev2012tasteweights} proposed an interactive recommender system that utilizes a three-column display. On the left side, users can adjust the weights assigned to items and content that contribute to the recommendation results, and they can cease adjusting once they are satisfied with the recommended items. A similar interface was created in \cite{schaffer2015hypothetical}, allowing users to add, delete, and re-rate items on the left side while observing the resulting changes in recommendations. In these studies, users are aware of the impacts of their interactions. However, when compared to our proposed method, the aforementioned works have several limitations: 1) Users are not provided with explanations for each individual recommendation, forcing them to guess the relationship between their actions and the recommendations, often leading to trials and errors. 2) The longer users engage with the recommender system, the more complex their accumulated behavior records become, making it increasingly challenging for users to make accurate assumptions and control the system. Our method distinguishes itself from previous works in that we employ machine learning approaches to generate counterfactual explanations for controllability. Additionally, our method is not restricted to a specific recommendation model; it can be applied to any recommendation model that relies on user behaviors for making recommendations.

\section{User Controllable Recommendation}
\label{sec:concepts}

In this section, we introduce the two types of controllable counterfactual explanations in Section \ref{sec:retro} and Section \ref{sec:pro}. Subsequently, we mathematically formulate the user-controllable recommendation framework in Section \ref{sec:notation}. Lastly, in Section \ref{sec:complexity_accuracy}, we discuss and define two fundamental aspects of user-controllable recommendation: the complexity of controllability and the accuracy of controllability.

\subsection{Counterfactual Retrospective Explanation}
\label{sec:retro}

Counterfactual retrospective explanations are derived from users' past behaviors and address the question: "If past behaviors were different, would this item still be recommended?" These explanations enable users to understand which historical actions influenced current recommendations. They take the following form:
\vspace{1em}
\begin{adjustwidth}{.5em}{1cm}
\fbox{\begin{minipage}{25em}
\footnotesize
\texttt{We recommend this item because you interacted with [item(s)]. Revoke these behaviors to stop its recommendation.}
\end{minipage}}
\end{adjustwidth}
\vspace{1em}

The term ``interacted with'' can be substituted with expressions such as ``liked,'' ``viewed,'' ``purchased,'' ``clicked,'' and others, depending on the specific form of interactions within the platform. Control via counterfactual retrospective explanation in recommendation is illustrated in the top-right part of Figure \ref{fig:overview}.

\subsection{Counterfactual Prospective Explanation}
\label{sec:pro}
Counterfactual prospective explanations are generated based on users' present behaviors. These explanations offer insights into future outcomes. When users engage in a specific behavior, prospective explanations ask the question: "If this behavior is carried out, how will future recommendations differ compared to not engaging in it?". They take the following form:

\vspace{1em}
\begin{adjustwidth}{.5em}{1cm}
\fbox{\begin{minipage}{25em}
\footnotesize
\texttt{With the current interaction, [item(s)] will be added to future recommendations. Revoke this behavior to prevent their recommendation.}
\end{minipage}}
\end{adjustwidth}
\vspace{1em}

Similarly, the choice of a specific interaction type depends on the real-world context. Control via counterfactual prospective explanation is illustrated in the bottom-right part of Figure \ref{fig:overview}.

\subsection{Definition of User Controllable Recommendation}
\label{sec:notation}
Technically, the proposed framework is applicable to any recommender system that generates recommendations based on users' historical behaviors. In this paper, we specifically introduce the controllable recommendation problem within the context of sequential recommendation. This choice is motivated by the fact that sequential recommender systems generate recommendations by considering user-item interactions in chronological order, and the time-aware nature of such systems aligns seamlessly with the concept of retrospective and prospective controllability.

Suppose a user set $\mathcal{U}=\{u_1, u_2, \cdots, u_m\}$ with $m$ users, and an item set $\mathcal{V}=\{v_1, v_2, \cdots, v_n\}$ with $n$ items. For each item $v_j$, $\bm{e_j}$ represents the related $d$-dimensional item embedding. A recommendation model learns the embedding matrix $\bm{E}\in \mathbb{R}^{n\times d}$ for all items.

For each user $u_i$ with last $T$ timestamps of the interaction history $\{v_i^t\mid t=1, \cdots, T\}$, $\bm{S_i}=[\bm{e_i^t} \mid t=1, \cdots, T]$ denotes the concatenated item embeddings. $T$ is a hyper parameter defined by the system, which indicates the maximum length of user history to be considered when making a recommendation. 

A recommender system predicts ranking score $r_{ij}$ for a user $u_i$ on an item $v_j$ based on $\bm{S_i}$, which can be generally represented as:
\begin{equation}
\label{eq:base}
    r_{ij} = f(v_j, \bm{S_i} \mid \Theta)
\end{equation}
where $\Theta$ are the model parameters. The function $f$ depends on the method used in the base recommendation model. In the rest of the paper, Eq. \eqref{eq:base} is simplified as $r_{ij} = f(v_j, \bm{S_i})$.

With the above ranking function, for each user $u_i$, the recommender system will generate a top-K recommendation, denoted as $\mathcal{R}(i, K)$. Under the the retrospective controllable recommendation scenario, if $u_i$ is not satisfied with some items in $\mathcal{R}(i, K)$, the system should allow the user to change their history $\bm{S_i}$ by removing certain previous actions, such that the recommendation results will be changed to the desired direction. 

Suppose the intervention is represented by a binary vector $\Delta_i \in \{0, 1\}^T$, which will be applied to $\bm{S_i}$ to form a counterfactual user history $\bm{S_i}^*$ such that:
\begin{equation}
    \bm{S_i}^*=\bm{S_i} \odot (\mathbb1 - \Delta_i)
\end{equation}
where $\odot$ stands for the element-wise product. Each item embedding $\bm{e}_i^t \in \bm{S_i}$ will be multiplied by the related $1-\delta_t$, where $\delta_t \in \Delta_i$ is the t$^{th}$ element in $\Delta_i$. When $\delta_t = 0$, the item embedding $\bm{e}_i^t$ at timestamp $t$ remains unchanged, which means the corresponding behavior of user $u_i$ at timestamp $t$ is preserved. When $\delta_t = 1$, $\bm{e}_i^t$ will become the padding embedding after multiplication, which means the related behavior is removed. After changing these behaviors, the base recommender will generate a new recommendation list $\mathcal{R}^*(i, K)$, which should better fit the user's expectation.

Under the prospective controllable recommendation scenario, the control option is more straight-forward. Suppose user $i$ interacts with a new item $v_{x}$ and generates a new user history embedding $[\bm{S_i}, \bm{e_x}]$. The user will be informed of the change in the future (i.e., the new items that will be added to the future recommendation). Users are allowed to decide whether this particular item, $v_x$, should be considered by the system or not in future recommendations.

\subsection{Complexity and Accuracy of Controllability}
\label{sec:complexity_accuracy}
Prior to the design and evaluation of a user-controllable recommendation model, it is crucial to establish the desired properties of an effective controllable method. In this paper, we propose two properties of controllability within the controllable recommendation scenario.

The first property is the complexity of controllability, which measures the cost for the user to control the recommendation results. Given that our framework relies on user behavior, complexity can be defined as the proportion of a user's past behaviors that must be eliminated in order to alter the recommendation results towards the intended direction. With the above notation, for a user $u_i$, when the change $\Delta_i$ is applied to the history embedding $\bm{S_i}$, the control complexity can be defined as:
\begin{equation}
\label{eq:complexity}
    \text{Complexity}~~=\frac{\|\Delta_i\|_0}{T},
\end{equation}
where $||\Delta_i||_0$ is the number of ones in $\Delta_i$, i.e., the number of removed behaviors of user $u_i$. We note that in reality, the length of some user histories may be smaller than $T$. In this case, it should be replaced by the real length of the user history. This applies to all the equations in the rest of the paper.

The second property is the accuracy of controllability. The control options offered by the system should accurately eliminate the undesired items from the recommendation list without inadvertently removing other relevant items. Suppose $u_i$ is unsatisfied with a subset of items in the recommendation list $\mathcal{R}_g(i, K) \subseteq \mathcal{R}(i, K)$, in an ideal situation, the item(s) removed from $\mathcal{R}(i, K)$ should only be the item(s) in $\mathcal{R}_g(i, K)$, which can be expressed as:
\begin{equation}
    \text{Accurate control} \Leftrightarrow \mathcal{R}(i, K) \setminus \mathcal{R}^*(i, K) = \mathcal{R}_g(i, K)
\end{equation}

However, a $100\%$ accurate control may be impossible to acquire. Thus, we further define the accuracy of the controllability as the Jaccard similarity (also known as Intersection over Union (IoU)) between the set of removed items and the set of undesired items. This can be formulated as follows:

\begin{equation}
\label{eq:accuracy}
    \text{Accuracy}~~= \frac{\mid \big( \mathcal{R}(i, K) \setminus \mathcal{R}^*(i, K)\big ) \cap \mathcal{R}_g(i, K) \mid }{\mid\big (\mathcal{R}(i, K) \setminus \mathcal{R}^*(i, K)\big)\cup \mathcal{R}_g(i, K)\mid}
\end{equation}

The accuracy of controllability should be maximized to ensure that only the undesired items are removed, while preserving the other items in the recommendation list. Simultaneously, the number of altered behaviors (i.e., the complexity) in the explanation should be minimized to enhance the comprehensibility of the explanation and facilitate the users in easily implementing the provided control options.

We note that these two properties not only guide the design of user controllable recommendation systems but are also highly suitable for quantitatively evaluating the controllabilities.

\section{\mbox{Generate Controllable Explanation}}
\label{sec:explgeneration}
In this section, we introduce the algorithms for generating the two types of explanations. 

\subsection{Generate Counterfactual Retrospective Explanation}
\label{sec:generate_retro}
To generate counterfactual retrospective explanations, we propose a constrained optimization problem that incorporates the previously defined complexity and accuracy of controllability.

The objective of the algorithm is to generate effective counterfactual retrospective explanations, ensuring that by revoking the past behaviors, the undesired items are removed from the recommendation list. This serves as a strict constraint in the optimization process. Meanwhile, as discussed in Section \ref{sec:complexity_accuracy}, the objective part of the optimization aims to minimize the complexity of the generated explanation while maximizing its accuracy. Considering the generation of an explanation to remove an item $v_j$ recommended to user $u_i$, and based on the mathematical definitions of complexity and accuracy, we formulate the optimization problem as follows:

\begin{equation}
\label{eq:originaloptimization}
    \begin{aligned}
    \minimize\limits_{\Delta_{i,j}}~~&\frac{\|\Delta_{i,j}\|_0}{T} - \gamma (\frac{\mid\big( \mathcal{R}(i, K) \setminus \mathcal{R}^*(i, K)\big ) \cap \{v_j\}\mid}{\mid\mathcal{R}(i, K) \setminus \mathcal{R}^*(i, K)\cup \{v_j\}\mid})\\
    &\text{s.t.,}~~v_j \notin \mathcal{R}^*(i, K)
\end{aligned}
\end{equation}
where $\gamma$ is a hyper-parameter to control the two terms on the same scale. Since $\frac{||\Delta_{i,j}||_0}{T} \sim ||\Delta_{i,j}||_0$ and $v_j$ is guaranteed to be removed as in the constraint, Eq.\eqref{eq:originaloptimization} is equivalent to:

\begin{equation}
\label{eq:firstoptimization}
    \begin{aligned}
    &\minimize\limits_{\Delta_{i,j}}~~\|\Delta_{i,j}\|_0 - \gamma \frac{1}{\mid\mathcal{R}(i, K) \setminus \mathcal{R}^*(i, K)\mid}\\
    &\text{s.t.,}~~v_j \notin \mathcal{R}^*(i, K) 
\end{aligned}
\end{equation}
In Eq. \eqref{eq:firstoptimization}, the second term is, in fact, maximizing the intersection between the original and counterfactual recommendation lists. Consequently, for the purpose of enhancing simplicity and feasibility of optimization, we reformulate the equation as follows:

\begin{equation}
\label{eq:simpleoptimization}
\begin{aligned}
&\minimize\limits_{\Delta_{i,j}}~~\|\Delta_{i,j}\|_0 - \gamma \sum_{v_q\in \mathcal{R}(u_i, K), v_q \neq v_j} \mathbb{I}(v_q, \mathcal{R}^*(i, K))\\
    &\text{s.t.,}~~v_j \notin \mathcal{R}^*(i, K) 
\end{aligned}
\end{equation}

where $\mathbb{I}(v_q, \mathcal{R}^*(i, K) )=1$ if $ v_q \in \mathcal{R}^*(i, K)$, otherwise $0$.

After solving $\Delta_{i,j}$, for $ \forall \delta_{t} \in \Delta_{i,j}$, if $\delta_{t} = 1$, the related interacted item $v_i^t$ will be included in the generated explanation set $\mathcal{E}_{i,j}$. However, searching for an optimal set of past behaviors is exponential. For a behavior set with size $x$, the searching space contains $\mathcal{O}(2^x)$ subsets. Therefore, inspired by a similar approach in \cite{goyal2019counterfactual}, we provide two options to approximate the optimization problem. One based on greedy search, and the other utilizing continuous relaxation.

\textbf{Greedy Search:} 
In the greedy search approach, we iteratively choose a single behavior to revoke at each time step, until the constraint in Eq. \eqref{eq:simpleoptimization} is satisfied, i.e., the accumulative effects remove item $v_j$ from the recommendation list $\mathcal{R}(i, K)$. At each step, we choose the behavior that fits both the constraint and objective in Eq. \eqref{eq:simpleoptimization} best. More specifically, instead of globally optimizing $\Delta_{i,j}$, we choose each item at one time by learning a one-hot vector $\bm{p}_{i, j} ^ {(r)}$ at each iterative step $r$. By removing the related item from user history, the ranking score of the target item $v_j$ should be as small as possible while the ranking scores of the other items in $\mathcal{R}(i, K)$ should be as large as possible. This strategy is guided by minimizing a heuristic function $h(\bm{p}_{i,j}^{(r)})$:

\begin{equation}
\label{eq:greedyopti}
\begin{aligned}
    &\minimize\limits_{\bm{p}_{i,j}^{(r)}}~~ h(\bm{p}_{i,j}^{(r)})=f(v_j, \bm{S_i}^{*(r)}) - \gamma_1 \sum_{v_q\in R(i, K), v_q\neq v_j} f(v_q, \bm{S}_i^{*(r)})\\
     &\text{s.t.,}~~\bm{p}_{i, j} ^ {(r)} = \{0, 1\}^T;~~ \|\bm{p}_{i, j} ^ {(r)}\|_1=1\\
    &\text{where}~~  \bm{S_i}^{*(r)} = \bm{S_i}^{*(r-1)} \odot (\mathbb{1} - \bm{p}_{i, j} ^ {(r)});~~ \bm{S_i}^{*(0)} = \bm{S_i}
\end{aligned}
\end{equation}

According to Eq. \eqref{eq:greedyopti}, we select one behavior from all the unchanged past behaviors to revoke at each step until $v_j$ is removed from the recommendation list, and return the change vector $\Delta_{i,j}=\sum_{r}\bm{p}_{i,j}^r$. If all the past behaviors have been changed but the item $v_j$ is still in the recommendation list, then the algorithm fails to generate retrospective explanation for this user-item pair. This process is illustrated in Algorithm \ref{alg}.

\begin{algorithm}[t]
\footnotesize
    \caption{Greedy Search}\label{alg}
    \begin{flushleft}
    \textbf{Input:} user $u_i$, target item $v_j$, history length $T$, user interaction history embedding $\bm{S_i}$, heuristic function $h$, length of recommendation list $K$. \\
    \textbf{Output:} the optimized explanation set $\mathcal{E}_{i,j}$.
    \end{flushleft}
    \begin{algorithmic}[1]
        \STATE Initialize explanation set
        $\mathcal{E}_{i,j} \leftarrow \{\}$
        \STATE $\bm{S_i}^*=\bm{S_i}$ 
        \WHILE{\textit{$v_j$ in the top-K recommendation} \textbf{AND} $|\mathcal{E}_{i,j}|<T$}
        \STATE $//$ Choose single behavior in a greedy manner.
        \STATE lowest\_heuristic $\leftarrow \infty$
        \FOR{\textit{each past behavior $\notin$ $\mathcal{E}_{i,j}$}}
        \STATE Generate one-hot vector $\bm{p}$
        \IF{$h(\bm{p})<\textit{lowest\_heuristic}$}
        \STATE update the best behavior according to $\bm{p}$
        \STATE update lowest heuristic 
        \ENDIF
        \ENDFOR
        \STATE Add the best behavior to $\mathcal{E}_{i,j}$
        \ENDWHILE
        \RETURN $\mathcal{E}_{i,j}$
    \end{algorithmic}
    \label{alg:model}
\end{algorithm}

\textbf{Continuous Relaxation:}
We provide another option to solve Eq. \eqref{eq:simpleoptimization} by transforming the original combinatorial problem into a continuous optimization problem. Since both the constraint part and the objective part are not differentiable, in order to learn the changing vector $\Delta_{i,j}$, we relax each component in Eq. \eqref{eq:simpleoptimization} into continuous space to provide approximate solutions. 

For the complexity objective, we relax the binary vector $\Delta_{i,j} \in \{0, 1\}^{T}$ as a continuous vector $\Delta_{i,j}^c \in [0, 1]^{T}$. $||\Delta_{i,j}^c||_1$ is used to approximate $||\Delta_{i,j}^c||_0$.

For the constraint, $v_j \notin \mathcal{R}^*(i, K)$. It is equivalent to the ranking score of $v_j$ is smaller than the ranking score of the item at the $K$th position of the recommendation list. Thus, we further relax it as a pair-wise loss between these two items, which is:
\begin{equation}
\begin{aligned}
    &R_1 = \max\big(0, \alpha_1 + f(v_j, \bm{S_i}^*)-f(v_K, \bm{S_i}^*)\big)\\
\end{aligned}
\end{equation}
where $v_K$ is the item at the $K$th position of the top-K recommendation list, which may be updated at each step.

For the accuracy objective $\gamma \sum_{v_q\in \mathcal{R}(i, K), v_q \neq v_j} \mathbb{I}(v_q, \mathcal{R}^*(i, K))$, we would like the ranking score of an item $v_q$ to remain as large as possible. Thus, we introduce another loss that measures the average ranking score of the items that shouldn't be removed, which is:

\begin{equation}
    R_2= \frac{1}{k-1} \sum_{v_q\in \mathcal{R}(i, K),v_q\neq v_j} f(v_q,\bm{S_i}^*)
\end{equation}

After the relaxations, the final equation of the optimization problem is reformulated as follows:
\begin{align}
\label{eq:relaxed}
    \begin{aligned}
        &\underset{\Delta_{i,j}^c}{\text{minimize}} ~\|\Delta_{i,j}^c \|_1 + \lambda (R_1 + \gamma_2 R_2) \\
    \end{aligned}
\end{align}

Eq. \eqref{eq:relaxed} is differentiable and can be solved by machine learning methods such as Stochastic Gradient Descent (SGD). After optimization, a threshold will be applied to $\Delta_{ij}^c$ to select the interacted items that will be included in the generated explanation $\mathcal{E}_{i,j}$. In this paper, we set the threshold to $0.5$. For continuous relaxation approach, the explanations should be generated only when they can successfully remove the target items. Thus, a post-checking process is required.

\textbf{Efficiency and Scalability Discussion:}
We note that in this framework, the complexity of the optimization space is linearly related to the fixed length $T$ of the user history and is irrelevant to the number of users and items in the system. This is attributed to the mechanism of counterfactual optimization. Empirically, when $T=100$ and the algorithm is executed on a single Nvidia A5000 GPU, it takes approximately $4$ seconds for the greedy search method to generate an explanation for a single recommendation, while the relaxation approach takes approximately $7$ seconds. The inference time does \textbf{not} encounter any scalability issues as the size of the dataset increases.

\subsection{Generate Counterfactual Prospective Explanation}
In comparison to searching for past behaviors that lead to a specific recommendation, it is more straightforward to determine the future consequences of the current behavior. This is due to the fact that when the recommender system is accessible, regardless of its transparency, the outcomes of any given current behavior are known.

Suppose a user $u_i$ interacts with an upcoming item $v_x$. The generation of counterfactual prospective explanations relies on assessing how the future recommendation would be altered by incorporating this particular user behavior into their history. Therefore, we update $\bm{S_i}$ with $\bm{S_i}^\prime = \bm{S_i} \cup \{v_x\}$. $\mathcal{R}(i,K)^\prime$ is the updated recommendation list in the future, which is generated based on $\bm{S_i}^\prime$.

Our user-controllable recommendation method looks for the items that are newly added to the recommendation list to form the counterfactual prospective explanation. For each item $v_j$:
\begin{equation}
    v_j \in \mathcal{E}_{{i,j}}, ~~~~~~~~ \text{if}~~ v_j \in \big( \mathcal{R}^\prime(i, K) \setminus \mathcal{R}(i, K) \big)
\end{equation}

If a user is dissatisfied with the change in the future recommendation list, they have the option to revoke the current interaction, thereby preventing the system from considering this behavior in the future. As the ``harmful'' interactions are eliminated dynamically, it is reasonable to assume that the recommender system can benefit from prospective controllability and deliver more precise recommendations in the future. In Section \ref{sec:proeval}, we conduct experiments to validate this assumption.

\section{Experiments}
\vspace{-5pt}
\label{sec:experiments}
\begin{table}
\centering
\caption{Statistics of the datasets.}
\label{tab:datasets}
\begin{adjustbox}{width=0.7\linewidth}
\begin{tabular}{ccccc}
\toprule
Dataset & \#User & \#Item & \#Interaction & Density \\ 
\cmidrule(lr){1-5}
ML-1M  & 6,040       & 3,416       & 987,540              & 0.031 \%        \\ 
Yelp    & 5,342       & 17,746       & 515,128              & 0.543 \%        \\ \bottomrule
\end{tabular}
\end{adjustbox}
\end{table}
\vspace{-10pt}

\begin{table}
\centering
\caption{Ranking accuracy of the base models}
\label{tab:base_acc}
\begin{adjustbox}{width=0.7\linewidth}
\begin{tabular}{ccccc}
\toprule
\multirow{2.5}{*}{Models} & \multicolumn{2}{c}{\textbf{ML-1M}} & \multicolumn{2}{c}{\textbf{Yelp}}  \\
\cmidrule(lr){2-3} \cmidrule(lr){4-5}
 & NDCG@10  & HT@10 & NDCG@10  & HT@10 \\
\cmidrule{1-5}
SASRec    & 0.1203   & 0.2265  & 0.0327  & 0.0518         \\
GRU4Rec   & 0.1150    & 0.2242  & 0.0145  & 0.0294         \\
\bottomrule
\end{tabular}
\end{adjustbox}
\end{table}
\vspace{-5pt}

\begin{table*}
\centering
\caption{Complexity and accuracy of controllability in retrospective evaluation. All numbers are percentages. UCR\_Relax and UCR\_Search are our proposed methods, which represent the relaxation approach and greedy search approach, respectively. All numbers are percentages.}
\label{tab:comacc}
\begin{adjustbox}{width=0.65\linewidth}
\begin{tabular}{ccccccccccccc}
\toprule
\multirow{4}{*}{SASRec} & \multicolumn{6}{c}{\textbf{ML-1M}} & \multicolumn{6}{c}{\textbf{Yelp}}  \\
\cmidrule(lr){2-7}\cmidrule(lr){8-13}
& \multicolumn{3}{c}{Complexity \%$\downarrow$}                                                    & \multicolumn{3}{c}{Accuracy \%$\uparrow$}                                 & \multicolumn{3}{c}{Complexity \%$\downarrow$}                                                    & \multicolumn{3}{c}{Accuracy \%$\uparrow$} \\
\cmidrule(lr){2-4}\cmidrule(lr){5-7} \cmidrule(lr){8-10} \cmidrule(lr){11-13}
& \multicolumn{1}{c}{@3}    & \multicolumn{1}{c}{@5}    & \multicolumn{1}{c}{@10}   & \multicolumn{1}{c}{@3}    & \multicolumn{1}{c}{@5}    & @10   & \multicolumn{1}{c}{@3}    & \multicolumn{1}{c}{@5}    & \multicolumn{1}{c}{@10}   & \multicolumn{1}{c}{@3}    & \multicolumn{1}{c}{@5}    & @10   \\
\cmidrule{1-13}

Dynamic\cite{schaffer2015hypothetical}    & 29.28   & 29.80  & 32.26  & 55.55  & 42.03  &  24.26 & 31.20  & 32.20 & 35.32 &52.17  & 39.03 & 22.61 \\
EFM\cite{zhang2014explicit}    & \underline{10.22}   & 11.88  & 13.29  & 58.21  & 46.16  & 33.30    & \underline{12.96} & 14.57 & 17.22 & 56.88 & 47.29 & \underline{33.29}\\
\cmidrule{2-13}
UCR\_Relax    & 12.16   & \underline{10.54}   & \underline{9.67}  & \underline{67.02}  & \underline{53.14}  & \underline{36.40}  & 16.22  & \underline{14.54} & \underline{14.02} & \underline{64.27} & \underline{50.70} & 32.26 \\
UCR\_Search    & \textbf{4.99}   & \textbf{4.58}  & \textbf{4.67}  & \textbf{76.79}  & \textbf{65.28}  & \textbf{48.02}  & \textbf{8.23}  & \textbf{7.92} & \textbf{7.88} & \textbf{78.94} & \textbf{68.02} & \textbf{49.52} \\
\cmidrule{1-13}
\multirow{4}{*}{GRU4Rec} & \multicolumn{6}{c}{\textbf{ML-1M}} & \multicolumn{6}{c}{\textbf{Yelp}}  \\
\cmidrule(lr){2-7}\cmidrule(lr){8-13}
& \multicolumn{3}{c}{Complexity \%$\downarrow$}                                                    & \multicolumn{3}{c}{Accuracy \%$\uparrow$}                                 & \multicolumn{3}{c}{Complexity \%$\downarrow$}                                                    & \multicolumn{3}{c}{Accuracy \%$\uparrow$} \\
\cmidrule(lr){2-4}\cmidrule(lr){5-7} \cmidrule(lr){8-10} \cmidrule(lr){11-13}
& \multicolumn{1}{c}{@3}    & \multicolumn{1}{c}{@5}    & \multicolumn{1}{c}{@10}   & \multicolumn{1}{c}{@3}    & \multicolumn{1}{c}{@5}    & @10   & \multicolumn{1}{c}{@3}    & \multicolumn{1}{c}{@5}    & \multicolumn{1}{c}{@10}   & \multicolumn{1}{c}{@3}    & \multicolumn{1}{c}{@5}    & @10   \\
\cmidrule{1-13}

Dynamic\cite{schaffer2015hypothetical}    & 26.89   & 29.67  & 32.49  & 60.61  & 41.52  & 25.27  & 46.47  & 49.51 & 48.83 & 57.92 & 40.72 & 24.87 \\
EFM\cite{zhang2014explicit}    & 17.35   & 15.48  & 17.14  & 62.33  & 45.59  & 29.54  & 33.19  & 32.78 & 34.11 & 61.83 & 45.65 & 27.33 \\
\cmidrule{2-13}
UCR\_Relax    & \underline{14.73}   & \underline{13.84}  & \underline{13.25}  & \underline{74.49}  & \underline{52.93}  & \underline{30.52}  & \textbf{17.12}  & \textbf{15.61} & \textbf{15.49}  & \underline{75.57} & \underline{55.57} & \underline{32.13} \\
UCR\_Search   & \textbf{5.51}   & \textbf{5.69}  & \textbf{6.40}  & \textbf{77.66}  & \textbf{59.38}  & \textbf{38.46}  & \underline{19.59}  & \underline{19.87} & \underline{20.39} & \textbf{77.38} & \textbf{59.95} & \textbf{37.33} \\
\bottomrule
\end{tabular}
\end{adjustbox}
\label{tab:}
\end{table*}

\begin{table}
\centering
\caption{Fidelity of the relaxation method. All numbers are percentages.}
\label{tab:fidelity_relaxation}
\begin{adjustbox}{width=0.75\linewidth}
\begin{tabular}{ccccccc}
\toprule
\multirow{2.5}{*}{Methods} & \multicolumn{3}{c}{\textbf{ML-1M}} & \multicolumn{3}{c}{\textbf{Yelp}}  \\
\cmidrule(lr){2-4} \cmidrule(lr){5-7}
 & @3 \%  & @5 \% & @10 \%  & @3 \% & @5 \%  & @10 \% \\
\cmidrule{1-7}
SASRec    & 62.97   & 59.76  & 55.14  & 52.37  & 49.06  & 46.55      \\
GRU4Rec    & 82.00   & 80.70  & 78.31  & 54.30  & 52.26  & 50.30      \\
\bottomrule
\end{tabular}
\end{adjustbox}
\end{table}

\subsection{Dataset Description}
The experiments are conducted on two widely used public datasets: 1) MovieLens-1M\footnote{\url{https://files.grouplens.org/datasets/movielens}} \cite{harper2015movielens}: containing movie ratings from the MovieLens website. It has approximately $1$ million ratings; 2) Yelp\footnote{\url{https://www.yelp.com/dataset}}: containing users' interactions and ratings on various businesses. For the Yelp dataset, following previous works \cite{zhang2014explicit, wang2018explainable}, we remove the users with less than $20$ interactions and items with less than $10$ interactions. Table \ref{tab:datasets} shows the overall statistics of the two datasets.

\subsection{Base Recommenders}
We choose two popular sequential recommendation models as the base models:
\begin{itemize}
    \item GRU4Rec\cite{hidasi2015session}: It applies Recurrent Neural Network (RNN) with Gated Recurrent Units (GRU) to model user click sequences for session-based recommendation.
    \item SASRec\cite{kang2018self}: It uses the self-attention mechanism to recommend next items.
\end{itemize}

\subsection{Implementation Details}
\textbf{Training Base Models:} When training these two models, we follow the same setup as in their original papers. For GRU4Rec, we apply $1$ layer of GRU units. For SASRec, we apply $2$ self-attention blocks. The item embedding size is $100$ for both models. For both datasets, we set the maximum length $T$ of user sequences to $100$, which means only the last $100$ interactions in the user history are taken into consideration when making predictions. 

We use the same training strategy when training these two models on the two datasets. First, we follow the generic leave-one-out strategy \cite{bai2019ctrec, zamani2020learning} to split the last item in the user sequences as the test data. We note that besides the test item, we also remove a certain number of last interacted items from the user history for prospective evaluation purpose, which will be introduced in Section \ref{sec:proeval}. This doesn't affect the training of the base models. During training, we use Adam optimizer\cite{kingma2014adam} with $0.001$ learning rate. The batch size is $128$. We set dropout rate to $0.2$ to prevent overfitting. Then, we train the model until they get the highest validation performance. We report the performance of the trained base models in Table \ref{tab:base_acc}. We note that the ranking accuracy of the base models are not the focus of this work. After training, the learned parameters in the recommendation models will be fixed for evaluating the controllability.

\textbf{Generating Retrospective Explanations:}
For the relaxation approach in Eq.\eqref{eq:relaxed}, we set the hype-parameter $\gamma_2=1$, and $\lambda=10$. We also set the $\gamma_1=1$ in the greedy approach in Eq. \eqref{eq:greedyopti}. These parameters have varies effects on the generated explanations, which are shown in the ablation study in Section \ref{sec:retroeval}. For solving the relaxation optimization problem, we use Adam optimizer with learning rate equals to $0.01$ and learn the change vector for $500$ steps.

\textbf{Generating Prospective Explanations:}
Given that prospective explanations aim to inform users about the potential outcomes of their ongoing interaction, we employ the next one item following the training sequence to simulate this process. To be precise, for each user, we append the next item to the end of the training sequence and observe the resulting ``future'' changes. All evaluations are conducted based on this procedure, which we elaborate on in Section \ref{sec:proeval}.

\subsection{Retrospective Evaluation}
\label{sec:retroeval}
We randomly sample $1000$ users as the test set and generate retrospective explanations for them. When evaluating the controllability, we revoke the interactions included in the generated controllable explanations by changing the related item embeddings to the padding embedding in the user interaction sequences.

Then, we evaluate the controllability according to the complexity and accuracy as introduced in Section \ref{sec:complexity_accuracy}. 

\begin{figure*}[t]
\mbox{
\centering
\hspace{40pt}
    \subfigure[Influence of $\gamma_1$]{
        \includegraphics[width=0.2\textwidth]{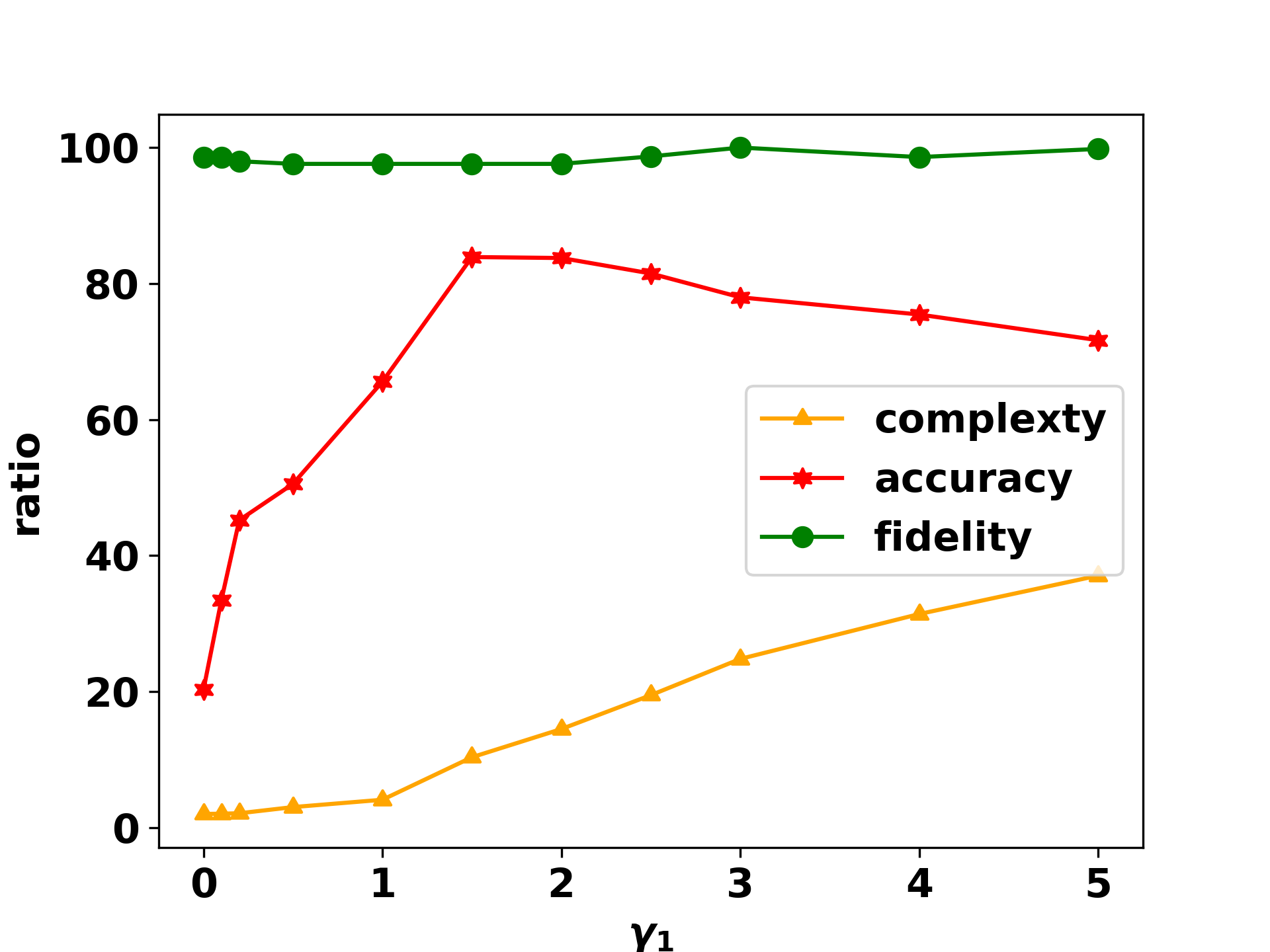}}
    \subfigure[Influence of $\lambda$]{
        \includegraphics[width=0.2\textwidth]{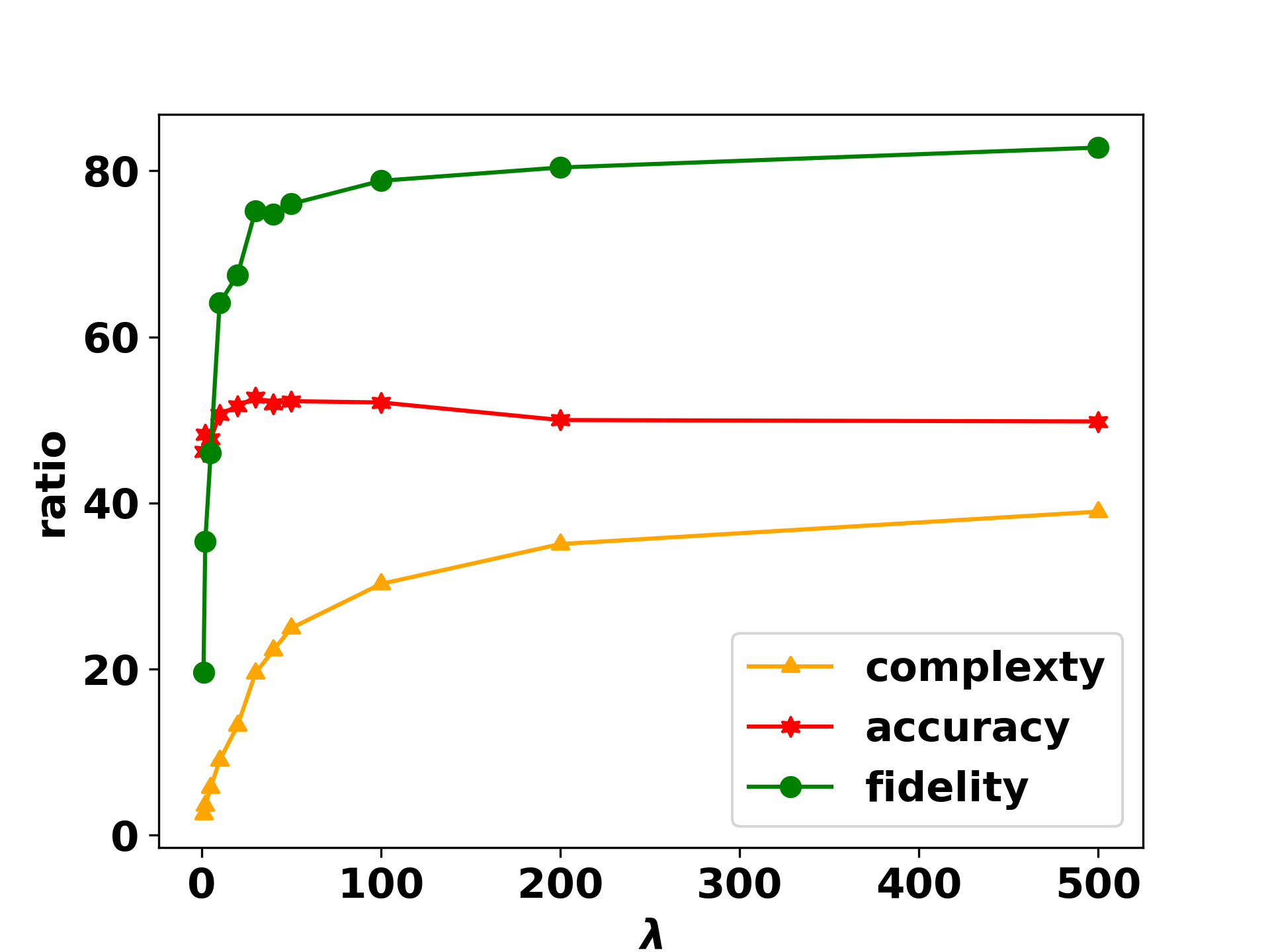}}
    \subfigure[Influence of $\gamma_2$]{
        \includegraphics[width=0.2\textwidth]{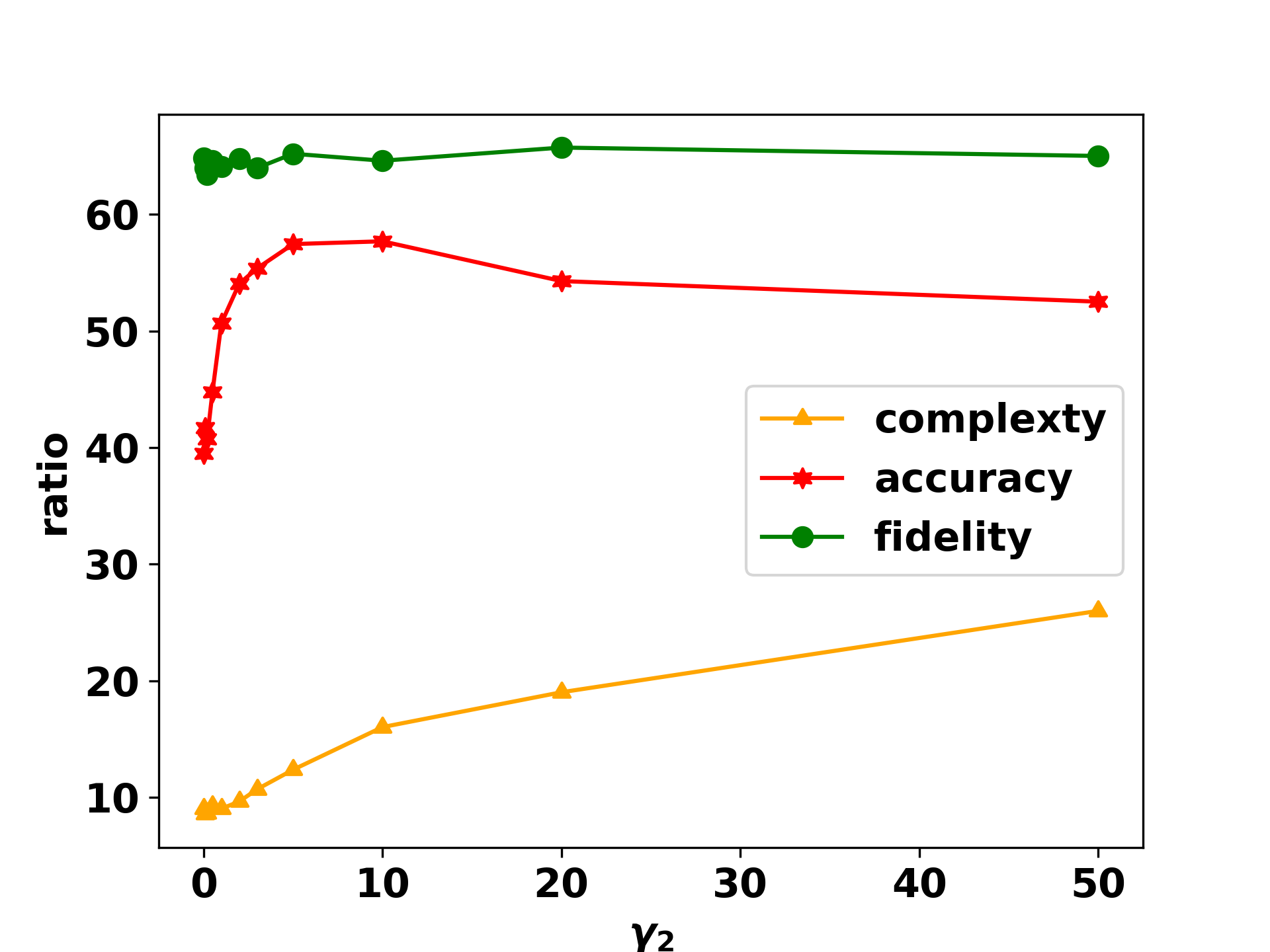}}
    \subfigure[Fid./Comp. vs. item position]{
        \includegraphics[width=0.2\textwidth]{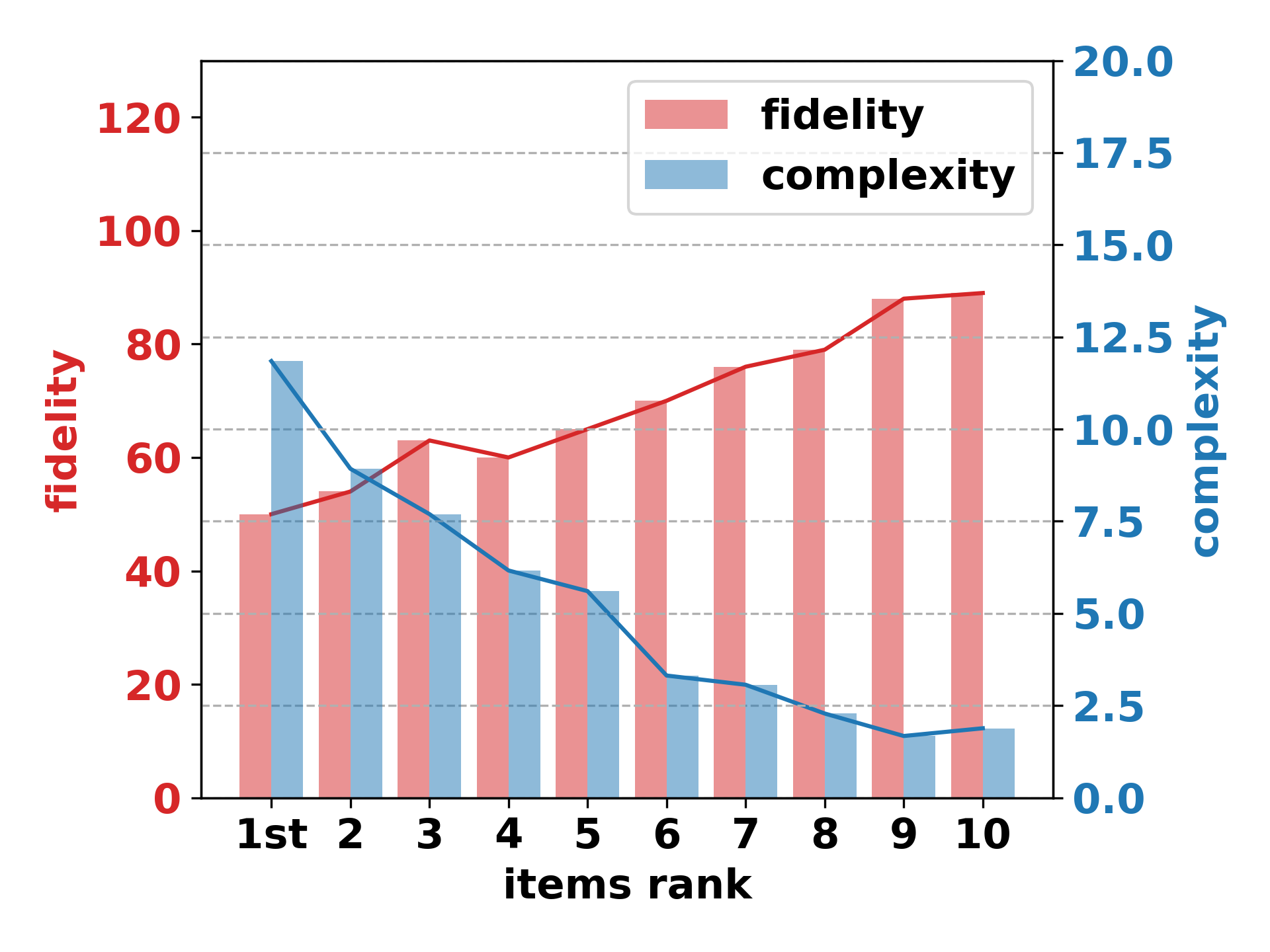}}
}
\vspace{-10pt}
\caption{Influence of the hyper-parameters according to the experiments on SASRec with MovieLens dataset. (a) Influence of $\gamma_1$ in greedy approach. (b) Influence of $\lambda$ in relaxation approach. (c) Influence of $\gamma_2$ in relaxation approach. (d) Fidelity and Complexity to remove items at different
ranking positions.}
\label{fig:ablation}
\vspace{-5pt}
\end{figure*}
\textbf{Comparable Baselines:}
To the best of our knowledge, there is no such paper that proposes an optimization algorithm to control the recommender system based on the user behaviors. Thus, we adjust two existing works in controllable recommendation and explainable recommendation to make them comparable: 1) \textbf{Dynamic Feedback} \cite{schaffer2015hypothetical}: It proposes a manual process to provide controllability, allowing the users to randomly remove the past behaviors until they are satisfied with the recommendation list. We create a random search algorithm to simulate this manual process; 2) \textbf{EFM} \cite{zhang2014explicit}: An explainable recommendation framework grounded in the concept of matching, a common approach in explainable recommendation \cite{wang2018explainable, chen2020try}. If an item is recommended, similar items in the user's history may be the reason for that recommendation. In each step, we remove the best-aligned item (according to the learned latent embeddings) until the target item is removed.

\textbf{Quantitative Analysis:}
In Table \ref{tab:comacc}, we evaluate the complexity and accuracy of the controllability for each baseline. First, we observe that both the greedy search and continuous relaxation methods provide significantly better controllability than the other methods. On average, based on the SASRec model, the relaxation method outperforms the second-best baseline by $2.20\%$ in terms of complexity and by $9.45\%$ in terms of accuracy. The greedy search method outperforms the second best baseline by $52.31\%$ in terms of complexity and by $41.49\%$ in terms of accuracy. While based on GRU4Rec model, the relaxation method outperforms the second-best baseline by $33.96\%$ in terms of complexity and by $16.74\%$ in terms of accuracy. The greedy search method outperforms the second-best baseline by $52.45\%$ in terms of complexity and by $29.68\%$ in terms of accuracy. It is worth noting that complexity is defined as the ratio between the number of items included in a generated explanation and the length of the user sequence. On average, the number of generated items is $3.48$ for the searching method and $5.66$ for the relaxation method. Although the relaxation and searching approaches produce comparable results when generating controllable explanations for the Yelp dataset based on GRU4Rec, according to all other results, the searching approach is generally the better option. Furthermore, the relaxation approach is unable to generate sufficient explanations for all the recommended items, for which we report the fidelity (i.e., the percentage of successful generation) in Table \ref{tab:fidelity_relaxation}. Therefore, we suggest always starting with the greedy search method when applying the proposed framework to new datasets or models.

Next, as shown in Figure \ref{fig:ablation}, we conduct ablation studies to verify the influences of $\gamma_1$ in the greedy search method, as well as $\lambda$ and $\gamma_2$ in the continuous relaxation method. First, Figure \ref{fig:ablation} (a) shows the influence of $\gamma_1$ in the greedy search method. When $\gamma_1$ is small, the control accuracy is highly positively related to it. However, when $\gamma_1>1$, the complexity increases dramatically. Therefore, choosing $\gamma_1=1$ is the optimal option. For the relaxation approach, as shown in Figure \ref{fig:ablation} (b), the $\lambda$ controls the trade-off between fidelity and complexity. When $\lambda$ increases, more effective explanations can be generated, but at the same time, the complexity also increases. Figure \ref{fig:ablation} (c) shows the influence of $\gamma_2$ on the relaxation method. It is very similar to the influence of $\gamma_1$ in the searching approach. When $\gamma_2$ is small, the control accuracy improves as $\gamma_2$ increases. However, when $\gamma_2>10$, choosing a larger $\gamma_2$ starts to negatively impact the model's performance. Furthermore, the complexity of controllability also steadily increases as $\gamma_2$ increases. 

Moreover, Figure \ref{fig:ablation} (d) displays the fidelity and complexity of the generated controllable explanations using the relaxation method, with respect to removing items at different positions in the recommendation list. It is evident that removing items ranked higher in the recommendation list is more challenging. This aligns with the notion that higher-ranked items have stronger reasons for their recommendation, thus more behaviors have to be revoked for the removing.

\vspace{-5pt}
\subsection{Prospective Evaluation}
\label{sec:proeval}
\begin{figure}[t]
    \centering
    \includegraphics[width=0.9\linewidth]{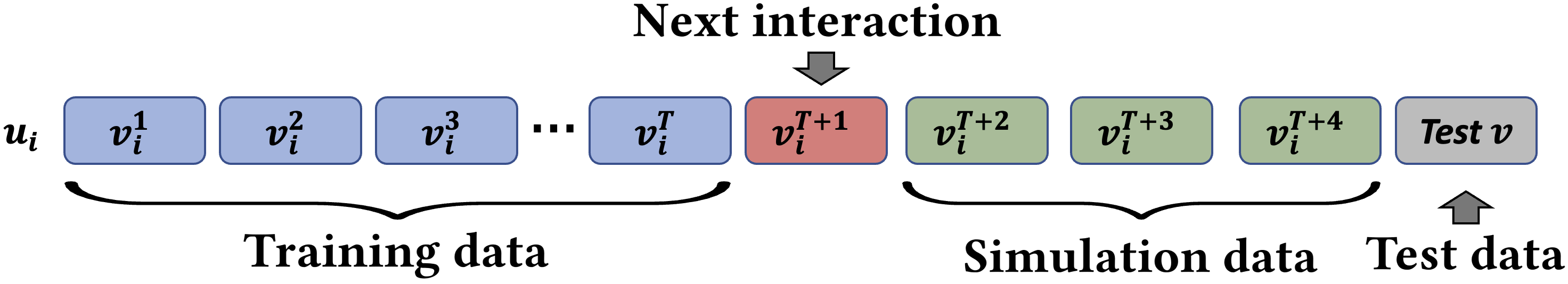}
    \vspace{-10pt}
    \caption{Data splitting in prospective evaluation.}
    \label{fig:data_split}
\vspace{-15pt}
\end{figure}

With prospective controllabilty, users have the ability to revoke their current behavior if they are dissatisfied with the ongoing changes, in hopes of receiving better recommendations in the future. In this section, we simulate this real-world scenario by introducing a novel data splitting strategy to validate this assumption. As illustrated in Figure \ref{fig:data_split}, the basic idea is as follows: in addition to excluding the test item, we also omit the last $M$ items in the user sequences as a "simulation set" that will not be observed during training. When a user interacts with a new item (i.e., adds the next interaction to the original user sequence), we examine the changes between the new recommendation list and the original recommendation list. If none of the newly added items are present in the simulation set, we can reasonably assume that in a real-world scenario, the user may not prefer this prospective change. Consequently, they would revoke this behavior. Then, for all user sequences that satisfy this condition, we test whether removing the current behavior from the user sequences will improve the accuracy on the test data.

During the experiments, we set $M=20$. We note that we concatenate the original user sequence, regardless of whether the current item is removed or not, with the simulation subsequence when making predictions. This is because we would like to test the long-term benefits of the prospective control option, rather than relying on short-term trivial observations. As shown in Table \ref{tab:prospective_acc}, we observe some intriguing results. When we remove the next interaction for \textbf{all} users and compare it to making predictions with the unchanged user history, the ranking accuracy drops significantly. This outcome is expected because removing an item from the user sequence results in the loss of valuable information. However, if we only remove the next interacted items for the users selected by the aforementioned simulation process, the overall ranking accuracy for this subset of users actually improves. This proves that, in addition to enhancing the transparency of the recommender system, providing system users with prospective controllability significantly increases the likelihood of receiving more accurate recommendations in the future.

\begin{table}
\centering
\caption{Change of ranking accuracy}
\label{tab:prospective_acc}
\begin{adjustbox}{width=0.8\linewidth}
\begin{tabular}{ccccc}
\toprule
\multirow{2.5}{*}{SASRec} & \multicolumn{2}{c}{\textbf{ML-1M}} & \multicolumn{2}{c}{\textbf{Yelp}}  \\
\cmidrule(lr){2-3} \cmidrule(lr){4-5}
 & NDCG@10  & HT@10 & NDCG@10  & HT@10 \\
\cmidrule{1-5}
All users    & 0.1203   & 0.2265  & 0.0327  & 0.0518         \\
All users (revoke)    & 0.0987 ($\downarrow$)   & 0.1905 ($\downarrow$)  & 0.0244 ($\downarrow$) & 0.0423 ($\downarrow$)        \\
\cmidrule(lr){2-5}
Target users   & 0.1078   & 0.1995  & 0.0301  & 0.0479         \\
Target users (revoke)  & 0.1118 ($\uparrow$)   & 0.2104 ($\uparrow$) & 0.0322 ($\uparrow$)  & 0.0496($\uparrow$)        \\
\cmidrule{1-5}
\multirow{2.5}{*}{GRU4Rec} & \multicolumn{2}{c}{\textbf{ML-1M}} & \multicolumn{2}{c}{\textbf{Yelp}}  \\
\cmidrule(lr){2-3} \cmidrule(lr){4-5}
 & NDCG@10  & HT@10 & NDCG@10  & HT@10 \\
\cmidrule{1-5}
All users    & 0.1150   & 0.2242  & 0.0145  & 0.0294         \\
All users (revoke)    & 0.0973 ($\downarrow$)   & 0.1914 ($\downarrow$) & 0.0136 ($\downarrow$) & 0.0278 ($\downarrow$)        \\
\cmidrule(lr){2-5}
Target users   & 0.1070   & 0.2074  & 0.0138  & 0.0283         \\
Target users (revoke)  & 0.1082 ($\uparrow$)   & 0.2094 ($\uparrow$) & 0.0140 ($\uparrow$) & 0.0283 (-)        \\
\bottomrule
\end{tabular}
\end{adjustbox}
\end{table}
\vspace{-10pt}

\section{Conclusion and Future Work}
In this paper, we present a framework that integrates explainability and controllability for user-controllable recommendation. By incorporating counterfactual retrospective and prospective explanations, our proposed framework enables users to control over the recommender system based on the user behaviors. As a first attempt, we introduce two fundamental properties of controllability: the complexity of controllability and the accuracy of controllability. These properties serve as general evaluation metrics for assessing the user-controllable recommendation problem. Controllability is important not only for recommendation, but also for many other AI sub-fields. In the future, we expect the exploration of this method beyond recommendation systems, extending its application to diverse AI systems, such as vision-based and language-based intelligent systems.\\

\textbf{Acknowledgement} This work was supported in part by NSF IIS
1910154, 2007907, 2046457, 2127918. Any opinions, findings, conclusions or recommendations expressed in this material are those of the authors and do not necessarily reflect those of the sponsors.

\bibliography{main}

\begin{thebibliography}{10}

\bibitem{Amatriain2009rateitagain}
Xavier Amatriain, Josep Pujol, Nava Tintarev, and Nuria Oliver, `Rate it again:
  Increasing recommendation accuracy by user re-rating', pp. 173--180, (01
  2009).

\bibitem{bai2019ctrec}
Ting Bai, Lixin Zou, Wayne~Xin Zhao, Pan Du, Weidong Liu, Jian-Yun Nie, and
  Ji-Rong Wen, `Ctrec: A long-short demands evolution model for continuous-time
  recommendation', in {\em Proceedings of the 42nd International ACM SIGIR
  Conference on Research and Development in Information Retrieval}, pp.
  675--684, (2019).

\bibitem{balog2019transparent}
Krisztian Balog, Filip Radlinski, and Shushan Arakelyan, `Transparent,
  scrutable and explainable user models for personalized recommendation', in
  {\em Proceedings of the 42nd International ACM SIGIR Conference on Research
  and Development in Information Retrieval}, pp. 265--274, (2019).

\bibitem{barbosa2020you}
Gabriel Diniz~Junqueira Barbosa and Simone Diniz~Junqueira Barbosa, `You should
  not control what you do not understand: the risks of controllability in ai',
  {\em Human Computer Interaction and Emerging Technologies: Adjunct
  Proceedings from},  231, (2020).

\bibitem{bostandjiev2012tasteweights}
Svetlin Bostandjiev, John O'Donovan, and Tobias H{\"o}llerer, `Tasteweights: a
  visual interactive hybrid recommender system', in {\em Proceedings of the
  sixth ACM conference on Recommender systems}, pp. 35--42, (2012).

\bibitem{chen2020try}
Tong Chen, Hongzhi Yin, Guanhua Ye, Zi~Huang, Yang Wang, and Meng Wang, `Try
  this instead: Personalized and interpretable substitute recommendation', in
  {\em Proceedings of the 43rd International ACM SIGIR Conference on Research
  and Development in Information Retrieval}, pp. 891--900, (2020).

\bibitem{ge2022survey}
Yingqiang Ge, Shuchang Liu, Zuohui Fu, Juntao Tan, Zelong Li, Shuyuan Xu, Yunqi
  Li, Yikun Xian, and Yongfeng Zhang, `A survey on trustworthy recommender
  systems', {\em arXiv preprint arXiv:2207.12515}, (2022).

\bibitem{ge2022explainable}
Yingqiang Ge, Juntao Tan, Yan Zhu, Yinglong Xia, Jiebo Luo, Shuchang Liu,
  Zuohui Fu, Shijie Geng, Zelong Li, and Yongfeng Zhang, `Explainable fairness
  in recommendation', in {\em Proceedings of the 45th International ACM SIGIR
  Conference on Research and Development in Information Retrieval}, pp.
  681--691, (2022).

\bibitem{ghazimatin2020prince}
Azin Ghazimatin, Oana Balalau, Rishiraj Saha~Roy, and Gerhard Weikum, `Prince:
  Provider-side interpretability with counterfactual explanations in
  recommender systems', in {\em Proceedings of the 13th International
  Conference on Web Search and Data Mining}, pp. 196--204, (2020).

\bibitem{goyal2019counterfactual}
Yash Goyal, Ziyan Wu, Jan Ernst, Dhruv Batra, Devi Parikh, and Stefan Lee,
  `Counterfactual visual explanations', in {\em International Conference on
  Machine Learning}, pp. 2376--2384. PMLR, (2019).

\bibitem{harper2015movielens}
F~Maxwell Harper and Joseph~A Konstan, `The movielens datasets: History and
  context', {\em Acm transactions on interactive intelligent systems (tiis)},
  {\bf 5}(4),  1--19, (2015).

\bibitem{Harper2015putting}
F.~Maxwell Harper, Funing Xu, Harmanpreet Kaur, Kyle Condiff, Shuo Chang, and
  Loren Terveen, `Putting users in control of their recommendations', RecSys
  '15, p. 3–10, New York, NY, USA, (2015). Association for Computing
  Machinery.

\bibitem{hidasi2015session}
Bal{\'a}zs Hidasi, Alexandros Karatzoglou, Linas Baltrunas, and Domonkos Tikk,
  `Session-based recommendations with recurrent neural networks', {\em arXiv
  preprint arXiv:1511.06939}, (2015).

\bibitem{hijikata2012relation}
Yoshinori Hijikata, Yuki Kai, and Shogo Nishida, `The relation between user
  intervention and user satisfaction for information recommendation', in {\em
  Proceedings of the 27th Annual ACM Symposium on Applied Computing}, pp.
  2002--2007, (2012).

\bibitem{hu2008collaborativefiltering}
Yifan Hu, Yehuda Koren, and Chris Volinsky, `Collaborative filtering for
  implicit feedback datasets', in {\em 2008 Eighth IEEE International
  Conference on Data Mining}, pp. 263--272, (2008).

\bibitem{jannach2016usercontrol}
D.~Jannach, Sidra Naveed, and Michael Jugovac, `User control in recommender
  systems: Overview and interaction challenges', in {\em EC-Web}, (2016).

\bibitem{Jannach2019explanationsandcontrol}
Dietmar Jannach, Michael Jugovac, and Ingrid Nunes, `Explanations and user
  control in recommender systems', in {\em Proceedings of the 23rd
  International Workshop on Personalization and Recommendation on the Web and
  Beyond}, eds., Mirjam Augstein, Eelco Herder, Wolfgang Wörndl, and Enes
  Yigitbas, p.~31, New York, NY, (2019). ACM.

\bibitem{jugovac2017interacting}
Michael Jugovac and Dietmar Jannach, `Interacting with recommenders—overview
  and research directions', {\em ACM Transactions on Interactive Intelligent
  Systems (TiiS)}, {\bf 7}(3),  1--46, (2017).

\bibitem{kang2018self}
Wang-Cheng Kang and Julian McAuley, `Self-attentive sequential recommendation',
  in {\em 2018 IEEE international conference on data mining (ICDM)}, pp.
  197--206. IEEE, (2018).

\bibitem{kingma2014adam}
Diederik~P Kingma and Jimmy Ba, `Adam: A method for stochastic optimization',
  {\em arXiv preprint arXiv:1412.6980}, (2014).

\bibitem{knijnenburg2012inspectabilityandcontrol}
Bart~P. Knijnenburg, Svetlin Bostandjiev, John O'Donovan, and Alfred Kobsa,
  `Inspectability and control in social recommenders', in {\em Proceedings of
  the Sixth ACM Conference on Recommender Systems}, RecSys '12, p. 43–50.
  Association for Computing Machinery, (2012).

\bibitem{knijnenburg2011each}
Bart~P Knijnenburg, Niels~JM Reijmer, and Martijn~C Willemsen, `Each to his
  own: how different users call for different interaction methods in
  recommender systems', in {\em Proceedings of the fifth ACM conference on
  Recommender systems}, pp. 141--148, (2011).

\bibitem{lamche2014interactive}
B.~Lamche, Ummahan Adıgüzel, and W.~Wörndl, `Interactive explanations in
  mobile shopping recommender systems', {\em CEUR Workshop Proceedings}, {\bf
  1253},  14--21, (01 2014).

\bibitem{mothilal2020explaining}
Ramaravind~K Mothilal, Amit Sharma, and Chenhao Tan, `Explaining machine
  learning classifiers through diverse counterfactual explanations', in {\em
  Proceedings of the 2020 conference on fairness, accountability, and
  transparency}, pp. 607--617, (2020).

\bibitem{art17gdpr}
European Parliament and Council of~the European~Union.
\newblock General data protection regulation (gdpr), art. 17, right to erasure
  (‘right to be forgotten’), 2016.
\newblock \newline\url{https://gdpr.eu/article-17-right-to-be-forgotten/}.

\bibitem{schaffer2015hypothetical}
James Schaffer, Tobias Hollerer, and John O'Donovan, `Hypothetical
  recommendation: A study of interactive profile manipulation behavior for
  recommender systems', in {\em The Twenty-Eighth International Flairs
  Conference}, (2015).

\bibitem{stepin2021survey}
Ilia Stepin, Jose~M Alonso, Alejandro Catala, and Mart{\'\i}n
  Pereira-Fari{\~n}a, `A survey of contrastive and counterfactual explanation
  generation methods for explainable artificial intelligence', {\em IEEE
  Access}, {\bf 9},  11974--12001, (2021).

\bibitem{tan2022learning}
Juntao Tan, Shijie Geng, Zuohui Fu, Yingqiang Ge, Shuyuan Xu, Yunqi Li, and
  Yongfeng Zhang, `Learning and evaluating graph neural network explanations
  based on counterfactual and factual reasoning', in {\em Proceedings of the
  ACM Web Conference 2022}, pp. 1018--1027, (2022).

\bibitem{tan2021counterfactual}
Juntao Tan, Shuyuan Xu, Yingqiang Ge, Yunqi Li, Xu~Chen, and Yongfeng Zhang,
  `Counterfactual explainable recommendation', in {\em Proceedings of the 30th
  ACM International Conference on Information \& Knowledge Management}, pp.
  1784--1793, (2021).

\bibitem{tan2023explainablefold}
Juntao Tan and Yongfeng Zhang, `Explainablefold: Understanding alphafold
  prediction with explainable ai', {\em arXiv preprint arXiv:2301.11765},
  (2023).

\bibitem{wang2018explainable}
Nan Wang, Hongning Wang, Yiling Jia, and Yue Yin, `Explainable recommendation
  via multi-task learning in opinionated text data', in {\em The 41st
  International ACM SIGIR Conference on Research \& Development in Information
  Retrieval}, pp. 165--174, (2018).

\bibitem{wasinger2013scrutable}
Rainer Wasinger, James Wallbank, Luiz Pizzato, Judy Kay, Bob Kummerfeld,
  Matthias B{\"o}hmer, and Antonio Kr{\"u}ger, `Scrutable user models and
  personalised item recommendation in mobile lifestyle applications', in {\em
  International Conference on User Modeling, Adaptation, and Personalization},
  pp. 77--88. Springer, (2013).

\bibitem{xiao2007commerce}
Bo~Xiao and Izak Benbasat, `E-commerce product recommendation agents: Use,
  characteristics, and impact', {\em MIS quarterly},  137--209, (2007).

\bibitem{zamani2020learning}
Hamed Zamani and W~Bruce Croft, `Learning a joint search and recommendation
  model from user-item interactions', in {\em Proceedings of the 13th
  International Conference on Web Search and Data Mining}, pp. 717--725,
  (2020).

\bibitem{zhang2020explainable}
Yongfeng Zhang and Xu~Chen, `Explainable recommendation: A survey and new
  perspectives', {\em Foundations and Trends in Information Retrieval}, (2020).

\bibitem{zhang2014explicit}
Yongfeng Zhang, Guokun Lai, Min Zhang, Yi~Zhang, Yiqun Liu, and Shaoping Ma,
  `Explicit factor models for explainable recommendation based on phrase-level
  sentiment analysis', in {\em Proceedings of the 37th international ACM SIGIR
  conference on Research \& development in information retrieval}, pp. 83--92,
  (2014).

\end{thebibliography}
\end{document}